\documentclass{article}
\usepackage[main, final]{neurips_2026}
\usepackage{multirow}
\usepackage{graphicx}
\usepackage{multicol}
\usepackage{amsfonts}
\usepackage{booktabs}
\usepackage{xcolor}
\usepackage{colortbl}
\usepackage{multirow}

\usepackage{amsmath}
\usepackage{array}
\usepackage{stfloats}
\usepackage{url}

\usepackage[utf8]{inputenc} 
\usepackage[T1]{fontenc}    
\usepackage{hyperref}       
\usepackage{nicefrac}       
\usepackage{microtype}      

\title{Phantomware: a Data Augmentation Framework for the Detection of Emerging Malware Campaigns}

\author{
  Antonino Rullo\\
  National Research Council \\ Institute for High Performance Computing and Networking (CNR-ICAR)\\
  via Pietro Bucci 8/9 C, 87036 Rende, Italy\\
  \texttt{antonino.rullo@cnr.it} \\
  }

\begin{document}
\maketitle

\begin{abstract}
Detecting unseen malware variants remains a major challenge as malware families continuously evolve. This issue is exacerbated in data-scarce scenarios such as emerging malware campaigns, where only a few samples are available to train robust detectors. A common strategy is to proactively generate malware variants and use them as augmentation data to expose models to a broader range of potential threats.
Existing approaches that generate functional malware variants through adversarial manipulation suffer from several important limitations. Generated samples must remain executable and preserve their malicious behavior, requiring costly validation through execution and analysis. Moreover, the adversarial objective provides limited control over the resulting augmentation dataset: it cannot guarantee the generation of sufficiently diverse variants, or even a single successful sample, making the effectiveness of the augmentation process inherently unpredictable. Finally, these approaches typically rely on a well-trained detector to guide generation, an assumption that may not hold in data-scarce scenarios such as early-stage malware campaigns.
We formally and empirically demonstrate these limitations and propose {\tt Phantomware}, a data augmentation framework designed for extreme data scarce scenarios. {\tt Phantomware} operates entirely in the representation space of a pretrained malware detector, avoiding the costly generation and validation of functional binaries. By exploiting latent-space geometry, it generates synthetic malware representations of a target family in a controlled manner, providing a constructive guarantee on the generation of diverse training samples. Extensive experiments on real malware families demonstrate that {\tt Phantomware} improves detection of unseen variants and provides robustness against both real and adversarially generated malware samples.
\end{abstract}

\maketitle

\section{Introduction}\label{sec:introduction}

Malicious software, commonly known as {\em malware}, has existed for decades and continues to represent one of the most persistent threats in modern computing systems. With the widespread adoption of the Internet and the growing interconnectedness of devices, the speed and scale at which malware can propagate have increased dramatically. Moreover, the evolution of artificial intelligence has enabled adversaries to deploy malicious code rapidly, diversify their attack vectors, and coordinate large-scale campaigns with unprecedented efficiency.

A major challenge in malware detection arises from the continual creation of malware variants. These variants are often generated through obfuscation techniques specifically engineered to evade existing defenses. Common evasion strategies, such as instruction reordering \cite{lucas2021malware}, function reordering \cite{ling2024wolf}, insertion of NOP operations \cite{park2019generation,zhang2022semantics}, and other semantic-preserving transformations, allow attackers to produce syntactically distinct programs that nevertheless retain their malicious functionality~\cite{brezinski2023metamorphic}. Malware variants that share the same behavioral intent but differ in their syntactic representation collectively constitute a {\em malware family}. In practice, variants belonging to well-represented families (i.e., families with a large number of known samples) are comparatively easier to detect, because classifiers can learn richer and more stable malicious patterns from the abundant training data. Conversely, variants from small families pose a harder detection problem due to the scarcity of available samples from which to learn discriminative features.

\paragraph{Motivations}
To address this issue, a natural strategy is to proactively generate malware variants that can supplement the training set, anticipating the evolution of small families. However, generating functional malware variants is time consuming because error-prone and computationally expensive: variants must remain executable and preserve the original malicious behavior, which typically requires validation through dynamic analysis in sandbox environments. 

Malware generation can also be performed {\em adversarially}, that is, by creating variants specifically designed to evade a target detection system. The attack seeks functionality-preserving perturbations that moves the classifier's decision toward the benign class while maintaining the malicious behavior of the original sample. The resulting evasion samples can then be used to retrain the detector, initiating a continual arms race in which the model is iteratively adapted to emerging attack strategies. However, despite the considerable progress achieved by adversarial malware generation as an offensive technique, its actual contribution to cybersecurity remains largely unexplored. Indeed, state-of-the-art adversarial approaches have been evaluated exclusively in terms of their evasion capability, where success is measured by their ability to bypass existing malware detectors (e.g., VirusTotal), whereas their effectiveness as defensive tools for improving detector robustness and generalization has been underinvestigated.
Moreover, these approaches suffer from three critical limitations: {\em (i)} the process is time-consuming due to the need for functional correctness checks; {\em (ii)} when only a limited number of malware samples is available, the generation procedure may fail to produce a sufficiently large and/or diverse number of adversarial samples for effective finetuning; and {\em (iii)} it assumes the availability of a well-trained detector, which is unlikely in scarce-data scenarios, making the adversarial component practically ineffective when dealing with early-stage malware campaigns.

An alternative approach is to generate {\em representations} of malware variants, that is, synthetic malware embeddings within a learned latent domain rather than constructing functional executables. These generated embeddings do not correspond to real malware samples; instead, they serve as constructive artifacts that approximate the potential distribution of small malware families. By enriching the embedding space in this way, detectors can be trained on a more diverse set of representations and can expand their decision boundaries toward regions that real future variants may plausibly occupy, eliminating the need for costly functionality validation. 
However, existing approaches rely on an adversarial generation objective, which provides only limited control over the location of the generated samples, as their primary goal is to induce misclassification. Consequently, the generated samples typically concentrate around the detector's decision boundary, providing limited coverage of other potentially challenging regions of the latent space. Moreover, these approaches generally assume access to tens or hundreds of real samples \cite{kim2017malware,tekerek2022novel}, and their effectiveness in early-stage malware campaigns, where only a handful of samples are available, has not been demonstrated.

\paragraph{Contribution}
To fill these gaps, we propose {\tt Phantomware}, a geometry-guided embedding generation approach which synthesizes malware-like latent representations by perturbing representative benign embeddings along malware-discriminative directions. The method does not require a large corpus of malware data, and leverages the geometric structure of the benign manifold to generate hard synthetic malware embeddings that populate challenging regions of the feature space. Since {\tt Phantomware} operates entirely in the learned latent space, the generated samples correspond to plausible feature-space variations of malware rather than fully functional or semantically meaningful executable files. These synthetic representations are subsequently used to finetune the detector, enhancing its robustness against underrepresented malware families. 
The name Phantomware reflects the essence of the approach: the generated embeddings are “phantoms” -- latent representations with no corresponding physical counterpart.\footnote{Phantom is broader term for any illusion or manifestation with no physical reality.}

Our contributions can be summarized as follow:

\begin{itemize}
    \item We provide a systematic assessment of adversarial malware generation techniques as data augmentation strategies. Through both formal analysis and extensive empirical evaluation, we demonstrate that existing adversarial approaches cannot guarantee the generation of sufficiently diverse and representative synthetic samples, thereby limiting their effectiveness in improving detector robustness and generalization.
    
    \item We present {\tt Phantomware}, a geometry-guided data augmentation framework designed for improving the detection performance of pretrained malware detectors with minimal computational effort and minimal reliance on malware data, making it suitable for extreme data-scarce scenarios such as emerging malware campaigns. We demonstrate the effectiveness of {\tt Phantomware} through an extensive experimental evaluation on both real malware families and adversarially generated malware samples. We make the {\tt Phantomware} implementation publicly available in an open-source repository.\footnote{https://github.com/ninorullo/phantomware/}

\end{itemize}

The rest of the paper is organized as follows. In Section \ref{sec:RW} we discuss related work;  
in Section \ref{sec:adversarialLimitations} we formally demonstrate the limitations of adversarial generation approaches; in Section \ref{sec:framework} we describe the {\tt Phantomware} framework; in Section \ref{sec:experiments} we empirically demonstrate either the effectiveness of {\tt Phantomware} and the shortcomings of adversarial approaches through an extensive set of experiments; finally, in Section \ref{sec:conclusions} we conclude the paper.
\section{Related Work}\label{sec:RW}

Malware research has progressed along two complementary directions, often regarded as two sides of the same coin: {\bf malware detection} and {\bf malware variant generation}. Detection methods typically rely on supervised models trained on large datasets and fall into two categories: {\bf static analysis} \cite{feng2024dawngnn,gibert2024certified,kim2018multimodal,raff2017malware,raff2021classifying}, which is fast and lightweight but struggles with obfuscated samples, and {\bf dynamic analysis} \cite{gulmez2024xran,li2022novel,qiang2022efficient}, which handles obfuscation better but is slower and resource-intensive.

The second research direction focuses on generating variants to anticipate attackers and improve detector robustness. {\em Problem-space} approaches produce variants that preserve malicious behavior, often using adversarial strategies aimed at evading a target detector. {\em Feature-space} approaches operate directly in the embedding space, producing synthetic malware representations in place of functional executables.

\subsection{Problem-space approaches}

Problem-space approaches generate functional malware variants by modifying malware artifacts at different levels of abstraction, including source code, assembly code, and binary code, as well as by leveraging intermediate representations such as embedding spaces. 
A key challenge across these approaches is preserving malware functionality. In particular, methods operating on intermediate or surrogate representations, such as embeddings, must map the generated variants back into the original malware space to obtain executable samples while preserving their behavior. 

We can be broadly distinguish these approaches into adversarial and non-adversarial, depending on whether the generated variants are explicitly designed to evade detection.

\subsubsection{Non-adversarial methods}

At the source-code level, researchers have leveraged the semantic and syntactic code comprehension abilities of Large Language Models to automate variant creation. Botacin \cite{botacin2023gpthreats} show that GPT-3 can readily obfuscate code, for example by generating multiple alternative implementations of the same function. Building on this idea, Akil et al. introduce LLMalMorph \cite{akil2025llmalmorph}, a Codestral-22–based framework that extracts function-level characteristics and uses custom prompts and transformation rules to guide variant generation without requiring model finetuning.

At the assembly-code level, other methods generate variants by inserting new API-call sequences into the program~\cite{lu2022research}, or by replacing the system call dependency graph with a semantically equivalent alternative, producing syntactically different yet behaviorally identical malware samples~\cite{ming2017impeding}.

\subsubsection{Adversarial methods}

Recent studies \cite{gao2024new,gibert2023wolf,hu2022generating,zhong2023malfox} employ GAN-based frameworks to generate adversarial malware samples capable of evading black-box detectors. In \cite{gao2024new,hu2022generating,zhong2023malfox}, the generator takes both benign and malicious samples as input to produce adversarial variants. These variants, together with benign samples, are submitted to the black-box detector to obtain target labels. The labeled data is then used to train the GAN’s discriminator, allowing it to serve as an effective surrogate for the black-box detector. Samples that successfully evade this surrogate model are considered valid adversarial examples. 
In \cite{gibert2023wolf}, Gilbert et al. propose applying mutations directly in the embedding space. The generator aims to produce malware samples whose feature representations closely resemble those of benign executables, while the discriminator outputs a “benignness” score for each sample. The feature set includes byte-frequency distributions, imported libraries and functions, and ASCII strings extracted from the executable.

Grosse et al. \cite{grosse2017adversarial}, followed by Greco et al. \cite{greco2025evasion}, employ the Jacobian-based Saliency Map Attack (JSMA) as the core mechanism of their malware generation frameworks. Introduced by Papernot et al. \cite{papernot2016distillation}, JSMA is an adversarial crafting technique that leverages the model’s Jacobian matrix to identify the minimal set of input features whose modification is most likely to induce misclassification. 

In \cite{ebrahimi2024learning,zhan2025game,zhan2024enhancing,zhang2022semantics} the authors propose a reinforcement learning–based approach in which an agent learns to modify executables to evade detection. During training, the agent observes the current version of the malware and decides which modification to apply.

Ling et al. \cite{ling2024wolf} modify the control-flow graph (CFG) of the original malware by replacing CALL instructions with a JUMP to a relocated CALL, producing a semantically equivalent executable with a different CFG.

Kolosnjaji et al. \cite{kolosnjaji2018adversarial} and Lucas et al. \cite{lucas2021malware} adopt a gradient-based optimization approach in which manipulations are iteratively applied to the executable as long as they move in the malware direction. Then, the process terminates as soon as the sample is classified as goodware. 

The manipulation and perturbation techniques employed in the adversarial approaches discussed above are diverse. Beyond code reorganization \cite{ling2024wolf}, malware variants are also commonly generated through code insertion and code modification techniques.
In \cite{choi2019amvg,demetrio2021functionality,xu2025automatic}, the authors use a genetic algorithm to mutate malware code by inserting pass statements \cite{choi2019amvg}, API calls \cite{xu2025automatic}, or padding bytes \cite{demetrio2021functionality}. Each mutated sample is tested against a detector, and further mutations are applied until evasion is achieved.
In \cite{demetrio2021functionality,kolosnjaji2018adversarial,kreuk2018adversarial}, the authors consider the addition of padding bytes only, to guarantee that the original malware code remains untuched. 
In \cite{park2019generation,zhang2022semantics}, the authors propose to insert NOP operations iteratively until the obtained variant evades a given detector.
In \cite{ebrahimi2024learning,suciu2019exploring}, bytes are injected to the empty sections of the executable (e.g., file overlay and empty spaces between sections) that does not affect the execution logic.
In \cite{digregorio2024tarallo,hu2022generating,xu2025automatic,zhan2025game,zhong2023malfox}, the authors select a set of API calls to be injected among those that do not disrupt the original functionality of the program.
Lucas et al. \cite{lucas2021malware} apply simple transformations to the assembly code, such as in-place randomization or code displacement.

Rosenberg et al. \cite{rosenberg2020generating} emulate a black-box detector using a white-box model, to which an explainability technique is applied to identify the features that most strongly influence classification. These features are then modified to generate adversarial samples. The selected features, such as the total number of strings, average string length, Windows path count, and byte histograms, are chosen to preserve program semantics. Similarly, De Rose et al. propose OLIVANDER \cite{de2025olivander}, a counterfactual-based method which provides LIEF-based explanations for model predictions by suggesting the minimal changes to input examples to modify the original prediction.

Only a limited number of studies have investigated the effectiveness of adversarial malware generation techniques as a defensive mechanism through adversarial training. Existing evaluations \cite{galovic2021improving,lucas2023adversarial} rely on the same attack strategies to generate both training and test samples, leaving it unclear whether the resulting models can generalize to unseen attacks. To fill this gap, Zainab et al. \cite{zainab2026breaking} have explicitly investigated the applicability of adversarial training, revealing substantial variability in attack effectiveness and transferability across models and datasets, and showing that adversarial hardening, although beneficial in some cases, does not consistently protect against diverse attack strategies.

\subsection{Feature-space approaches}

Unlike behavior-preserving malware generation techniques, which focus on producing functional executables, feature-space approaches aim to construct a sufficiently large and diverse set of training data to support effective learning. Since machine learning models operate on vectorial representations, it is sufficient to generate synthetic malware embeddings that closely resemble those observed in practice, as they can still provide meaningful training signals for the model.
A common approach to generate such vectorial representations is the use of a GAN \cite{al2026improving,ashawa2025static,burks2019data,chen2021using,kim2017malware,rahman2024vision,tekerek2022novel}, which learns the underlying distribution of real malware embeddings and generate new samples that lie on the same data manifold. Malware embeddings are often obtained as images derived from the malware bytecode, with each pixel determined by the byte value.

In \cite{kim2017malware}, Kim et al. train an autoencoder on malware samples and then repurpose the decoder as the generator in a GAN. After training the GAN, the discriminator is employed as a malware detector.
Rahman et al. \cite{rahman2024vision} propose a hybrid data augmentation technique based on a deep convolutional GAN plus traditional image transformation methods.
In \cite{ashawa2025static,tekerek2022novel}, the authors focus on generating realistic malware variants by learning feature transformations between similar samples using a CycleGAN. Tekerek and Yapici \cite{tekerek2022novel} implement this by training a CycleGAN on intra-family samples to synthesize new variants. Ashawa et al. \cite{ashawa2025static} train a CycleGAN to convert a static binary image into an image derived from a memory dump of its execution, and vice versa. Note that while CycleGANs are typically used for image translation between two distinct domains, in this setting the domains correspond to different malware samples within the same family. 

Recently, Yang et al. \cite{yang2025malimgda} have shown that diffusion models can be adopted as data augmentation tools in the malware domain. They pre-train a diffusion model on existing malware image data to capture generic malicious textures, then they finetune this foundation model using images from the target family. Finally, images from the target family act as seeds for the finetuned model, which iteratively generates new variants. The newly generated images are continually added back into the target family's training set to serve as seeds for subsequent epochs.

In \cite{ikram2022malimg2022} and \cite{catak2021data} the authors use image-specific data augmentation techniques such as feature standardization, image flipping, shifting, and rotation, as well as the addition of Gaussian, Poisson, and Laplace noise as data augmentation approaches.

In \cite{benedetti2025days,pasqualatto2026malarn}, the authors propose the use of a pre-trained LLM to map malware instructions into a latent space. Then, a variational autoencoder \cite{benedetti2025days} and an adversarial
reconstruction network \cite{pasqualatto2026malarn} are used to generating realistic embedding variants of malware.

Finally, Al-Essa et al. \cite{al2026improving} propose deriving synthetic malware embeddings from adversarial examples rather than from real samples. While this strategy outperforms methods relying solely on adversarial samples, it remains less efficient than approaches operating entirely within the latent space. Indeed, generating such synthetic data still requires first crafting adversarial examples in the original malware space, which entails the computational and semantic constraints associated with malware manipulation before any latent-space processing can occur. Furthermore, like the other works discussed so far, Al-Essa et al. do not address emerging malware campaign scenarios, as their method still relies on thousands of original malware to generate the adversarial examples.

\subsection{Discussion}

Despite the advancements in (adversarial) functional malware generation, several significant limitations persist, particularly regarding their real-world applicability and defensive utility. A primary challenge stems from the nature of acceptable perturbations in the malware domain: unlike image classification, where changes are continuous, malware inputs consist of discrete features, greatly reducing the span of acceptable modifications. Moreover, perturbations that are syntactically valid in the problem space may be irrelevant to the detector. This challenge is further compounded by the requirement that any modification must preserve malicious functionality, often necessitating dynamic analysis in sandbox environments.

Beyond inefficiency, several other factors limit the effectiveness of these approaches for defensive purposes. First, many existing methods rely on iterative random transformations, however, greedily selecting effective manipulations tends to reduce the diversity of adversarial perturbation sequences. Second, adversarial training is only effective against the specific perturbation pattern fed to the model, and exhaustively generating adversarial examples with all possible patterns may be highly time consuming. Third, the availability of a well-trained detector cannot be assumed in data-scarce settings, rendering the adversarial component largely ineffective for small malware families. Finally, the common assumption that attackers apply only minimal syntactic perturbations is not always true. For example, a ransomware family may evolve from directly encrypting files to using process injection to achieve the same malicious objective. Although the malicious intent remains unchanged, the execution semantics differ substantially, producing a semantic rather than purely syntactic variant. Consequently, obfuscation-based generation strategies may fail to capture such variations.

Overall, these limitations make adversarial malware generation a constrained strategy for learning robust and generalizable detection models from limited malware data, since they cannot guarantee the generation of sufficiently diverse, representative, and semantically rich training samples. Moreover, existing research heavily focuses on evasion capability, measuring success solely by how effectively samples bypass current detectors (e.g., VirusTotal), while completely overlooking whether these adversarial approaches actually improve a detector's robustness and generalization.

On the other hand, feature-space approaches are generally more efficient than techniques that operate in the malware space, as they avoid the need for constraint-driven generation and computationally expensive validation procedures. However, existing methods remain fundamentally adversarial in nature, generating synthetic representations by perturbing real ones to cross the detector's decision boundary. As a result, the generated embeddings tend to concentrate in the vicinity of the boundary, leaving more distant regions of the latent malware manifold largely unexplored. 

Finally, we note that both problem-space and embedding-based approaches depend on the availability of a sufficiently large and representative set of real malware samples. For the former, the diversity of generated variants is constrained by that of the original malware instances, whereas embedding-based methods are limited by the latent distribution learned from the available training samples. Consequently, in data-scarce scenarios, both approaches struggle to generate sufficiently diverse and representative synthetic data, limiting their ability to improve detector robustness and generalization.

To address these gaps, we propose an embedding-based data augmentation framework capable of generating synthetic malware representations starting from the available goodware samples and from a limited number of malware instances, enabling the generation of representative augmentation data even in highly data-scarce scenarios. 
\section{A Formal Proof of the Limitations of Problem-space Adversarial Generation Approaches}\label{sec:adversarialLimitations}

Adversarial malware generation can be naturally formulated as a constrained optimization problem, where the objective is to identify a functionality-preserving malware variant that is misclassified by a target detector. Starting from an original malware sample, an adversarial attack iteratively explores the space of admissible perturbations by generating candidate variants, evaluating them against the target detector, and selecting the next perturbation according to its optimization strategy (e.g., gradient-based or heuristic search). Each optimization step constitutes one iteration of the attack. The effectiveness of this process therefore depends on both the existence of feasible adversarial solutions and the ability of the optimization algorithm to discover them within a finite number of iterations. In the following, we show that neither property can be guaranteed in general. 

Let $m$ be a malware sample, let $D:\mathcal{X}\rightarrow\{0,1\}$ be a malware detector, where $D(x)=1$ denotes the malware class and $D(x)=0$ the goodware class, and let $\mathcal{V}(m)\subseteq\mathcal{X}$ denote the set of all functionality-preserving variants of $m$. An adversarial generation algorithm $A(m,D)$ seeks a sample $m'\in\mathcal{V}(m)$ such that $D(m')=0$. Given a malware $m$ and a detector $D$, we define the feasible evasion set as
\[
\mathcal{E}(m,D)=\{m'\in\mathcal{V}(m)\mid D(m')=0\}
\]
We immediately observe that the existence of an adversarial sample is not guaranteed. Indeed, $\mathcal{V}(m)$ is a discrete and highly constrained search space, since every perturbation must preserve the original malicious functionality. Consequently, it is possible that the detector correctly classifies every functionality-preserving variant as malware, i.e.,
\[
\quad D(m')=1, \forall m'\in\mathcal{V}(m)
\]
which implies $\mathcal{E}(m,D)=\emptyset$. In this case, the underlying optimization problem admits no feasible solution, and therefore no adversarial generation algorithm can produce a successful evading sample.

Even when $\mathcal{E}(m,D)\neq\emptyset$, no finite iteration budget can guarantee that an iterative adversarial algorithm $A$ will find an evading sample. Let
\[
T(m,D,A)=\min\{t>0 \mid A_t(m,D)\in\mathcal{E}(m,D)\}
\]
denote the first iteration at which the attack generates a successful adversarial example, where $A_t$ is the state of the attack after $t$ iterations. Since $T(m,D,A)=\infty$ whenever $\mathcal{E}(m,D)=\emptyset$, and may be arbitrarily large even when $\mathcal{E}(m,D)\neq\emptyset$, no finite upper bound $B<\infty$ exists such that $T(m,D,A)\leq B$ for every malware sample $m$, detector $D$, and attack $A$. Therefore, there exists no adversarial generation algorithm that can guarantee for every detector--malware pair the production of even a single successful evading sample within a finite number of iterations.

In practice, the number of adversarial samples generated from a given malware instance may vary significantly across attack strategies and can, in some cases, be zero. On the other hand, a given attack strategy may generate multiple successful variants for some malware samples, whereas for others it may fail to generate any. This uncertainty is further exacerbated in data-scarce scenarios, as the pool of adversarial samples is inherently constrained by the number and diversity of available real malware instances. As a result, both the size and diversity of the resulting synthetic malware set are difficult to control, limiting the reliability of adversarial generation as a data augmentation strategy.
\section{The Phantomware Data Augmentation Pipeline}\label{sec:framework}

The limitations of existing data augmentation pipelines motivate the need for a non-adversarial approach that minimizes dependence on malware data to address the severe data scarcity characterizing emerging malware campaigns. To this end, we present {\tt Phantomware}, a data augmentation pipeline that operates entirely in the representation space of a pretrained malware detector.
Unlike previous approaches, {\tt Phantomware} eliminates the need for binary manipulation and validation, avoids adversarial optimization procedures, and operates effectively in highly data-scarce settings, requiring even a single malware sample to characterize malware behavior while leveraging the diversity of the goodware population to generate representative synthetic malware embeddings.
The proposed framework is model-independent as it can be applied to any machine learning-based malware detector. 

\subsection{Problem formulation}\label{sec:problemFormulation}

Let $T$ denote a target malware family, and $\mathcal{M}^{T}={\mathbf{x}_1,\ldots,\mathbf{x}_n}$ denote the corresponding set of malware samples.
Our objective is to augment the set $\mathcal{M}^{T}$ with a synthetic dataset $\widetilde{\mathcal{M}}^{T}=\tilde{\mathbf{x}}_1 \ldots \tilde{\mathbf{x}}_q$ that captures relevant characteristics of $T$ while providing additional information about potential future variants. The resulting augmented dataset is then used to finetune a pretrained malware detector, with the goal of improving its ability to correctly classify unseen variants of $T$. To achieve this, $\widetilde{\mathcal{M}}^{T}$ must provide sufficient diversity with respect to the available real samples in $\mathcal{M}^{T}$. 
A key requirement is that the augmentation procedure remains effective when $|\mathcal{M}^{T}|$ is small and therefore does not depend on the availability of a large corpus of target-family samples.  Thus, the goal is not simply to increase the number of training samples, but to enrich the representation of a small malware family with informative samples that are complementary to the available real data.

In the context of {\tt Phantomware}, we address this problem directly in the latent space of a pretrained malware detector. Rather than generating functional malware executables and subsequently processing them through the detector, we directly generate their latent vector representations, as these are the representations on which the machine learning model ultimately operates to perform classification. This avoids the constraints associated with generating, validating, and executing functional malware while allowing the augmentation process to directly target the representation space relevant to the detector.

The proposed framework consists of two complementary phases: {\em (i)} a geometry-guided embedding generation method that produces synthetic representations of the target family, and {\em (ii)} a lightweight finetuning procedure that incorporates both real and synthetic representations to adapt the detector to previously unseen variants.

\subsection{Geometry-guided embedding generation}\label{sec:embGen}

The core intuition behind {\tt Phantomware} is to generate synthetic malware embeddings that lie close to the benign manifold while preserving discriminative characteristics of the target malware family, thereby populating benign-adjacent regions of the latent space with representative malicious samples. This design is motivated by the observation that the most challenging threats are not those located just beyond the decision boundary, but those whose representations closely resemble benign software, occupying regions where malicious samples are sparse or entirely absent. 

Unlike existing adversarial malware generation approaches \cite{gao2024new,gibert2023wolf,hu2022generating,zhong2023malfox}, which perturb real malware samples with benign content (e.g., API calls, byte-frequency distributions, or ASCII strings) to induce detector evasion, our approach exploits the abundance of benign samples to characterize latent regions where future evasive malware may emerge. Rather than perturbing the few available malware samples from the target family, we generate synthetic embeddings by moving representative benign anchors along malware-discriminative directions. The resulting samples preserve the local geometry of the benign manifold while acquiring malware-specific characteristics, yielding hard training examples that enrich regions of the latent space largely unexplored by conventional adversarial methods. By explicitly populating these peripheral regions with synthetic malware representations, {\tt Phantomware} exposes the detector to a broader spectrum of challenging scenarios, ultimately improving its robustness against subtle, benign-looking malware variants.

\subsubsection{Goodware anchor selection}
We select the goodware anchor set as the union of two complementary subsets of embeddings: \textbf{boundary samples} and \textbf{frontier samples}. Boundary samples are benign points located close to malware instances in the latent space and therefore near the decision boundary, where classification uncertainty is highest. Frontier samples, on the other hand, are benign points located at the periphery of the benign distribution. These samples capture low-density peripheral regions underrepresented in the training data, and where evading, benign-looking malware may appear.
Together, boundary and frontier samples offer a diverse set of anchors from which realistic and challenging malware-like embeddings can be synthesized.

Let $\mathcal{G}$ denote the set of goodware embeddings, and $\mathcal{M}^T$ denote the set of malware embeddings of the target family $T$. To identify boundary samples, we define for each point $\mathbf{g}\in\mathcal{G}$ a proximity-based score $s_{\text{pb}}$ measuring its closeness to the $\mathcal{M}^T$ manifold:
\[
s_{\text{pb}}(\mathbf{g}) = \log \sum_{j=1}^{|\mathcal{M}^T|} \exp\left(-\|\mathbf{g} - \mathbf{m}_j\|_2\right)
\]

where $\mathbf{m}_j\in\mathcal{M}^T$. The negative distance is converted into a similarity measure through the exponential function, such that malware samples closer to $\mathbf{g}$ contribute more strongly to the resulting score. The logarithm provides a smooth aggregation of these contributions, yielding a soft approximation of the negative distance to the nearest target-family malware sample. Consequently, goodware embeddings with higher scores are located in regions that are closer to the target-family malware representations and are therefore more likely to correspond to ambiguous regions of the latent space.

The set of boundary samples $\mathcal{B}$ is defined as the top-$k$ goodware embeddings with the highest proximity score:
\[
\mathcal{B} = \arg\max_{\mathbf{g}\in\mathcal{G}}^k\left(s_{\text{pb}}(\mathbf{g})\right)
\]

To capture the outer structure of the benign manifold, we identify frontier samples using a PCA-based representation. Let $\{\mathbf{v}_j\}_{j=1}^{k}$ denote the top-$k$ principal components of $\mathcal{G}$, obtained from a PCA decomposition of the goodware embedding space. We define the projection score of each embedding $\mathbf{g}_i \in \mathcal{G}$ onto direction $\mathbf{v}_j$ as

\[
s_{i,j} = \mathbf{g}_i^\top \mathbf{v}_j
\]

For each principal direction $j$, we select frontier samples as the $l$ largest and smallest projections:

\[
\mathcal{F}_j^{+} = \left\{ \mathbf{g}_i \in \mathcal{G} \;\middle|\; s_{i,j} \text{ is among the top-}l \right\}
\]
\[
\mathcal{F}_j^{-} = \left\{ \mathbf{g}_i \in \mathcal{G} \;\middle|\; s_{i,j} \text{ is among the bottom-}l \right\}
\]

The frontier set is defined as

\[
\mathcal{F} = \bigcup_{j=1}^{k} \left( \mathcal{F}_j^{+} \cup \mathcal{F}_j^{-} \right)
\]

The final goodware anchor pool $\mathcal{G^{\text{A}}} = \mathcal{B} \cup \mathcal{F}$ is obtained by combining boundary and frontier samples, thereby covering both decision-interface regions and the outer geometry of the benign manifold.

\subsubsection{Synthetic embedding generation}\label{sec:emb-generation}

To guide the synthesis process, we first identify the latent directions associated with malicious behavior by computing a malware-discriminative displacement vector that captures the differences between the malware and goodware embedding distributions. For each latent dimension $i$, we compute the difference between the corresponding class centroids:

\[
d_i = \mu_{M,i} - \mu_{G,i}
\]

where $\mu_{M}$ and $\mu_{G}$ are the mean embeddings of malware and goodware samples that belong to $\mathcal{M}^T$ and $\mathcal{G^{\text{A}}}$, respectively. The sign of $d_i$ encodes the direction of the shift toward the malware distribution, while its magnitude indicates how the latent dimension changes between the two classes.

To identify the dimensions that are most informative for distinguishing malware from goodware, we compute a Fisher-inspired discriminative score:

\[
F_i =
\frac{|d_i|}
{\sigma_{M,i}+\sigma_{G,i}+\epsilon}
\]

where $\sigma_{M,i}$ and $\sigma_{G,i}$ are the standard deviations of the $i$-th latent dimension for malware and goodware embeddings, respectively.
This score assigns higher importance to dimensions that exhibit a large separation between malware and goodware while having low intra-class variability.

The resulting scores are normalized and sharpened to obtain a continuous importance vector:

\[
w_i =
\left(
\frac{F_i}{\max_j F_j+\epsilon}
\right)
\]

Finally, the malware-discriminative direction is obtained by weighting the signed displacement vector:

\[
\Delta_i = d_i \cdot w_i 
\]

$\Delta$ preserves both the importance and the direction of each latent dimension, allowing the generation process to move benign embeddings toward the malware manifold while maintaining the intrinsic geometry of the latent space.

Given a benign anchor embedding $\mathbf{g}\in\mathcal{G^{\text{A}}}$, a synthetic malware embedding $\tilde{\mathbf{m}}$ is generated as:

\[
\tilde{\mathbf{m}} = \mathbf{g} + \alpha \Delta 
\]

where $\alpha$ controls the magnitude of the displacement. By varying $\alpha$, the procedure can generate a set of synthetic embeddings spanning different regions between the benign manifold and malware regions of the latent space. For the generation of each single $\tilde{\mathbf{m}}$, $\alpha$ is randomly sampled from a uniform distribution between 0.1 and 1.

This procedure generates hard synthetic samples that populate both the vicinity of the benign manifold and regions beyond the decision boundary. Notably, the approach ensures data generation even when only a single malware sample is available, as this suffices to extract malware-discriminative features that guide the perturbation of benign representations, making the method naturally suited to extreme data-scarce scenarios. In practice, the generation process provides a constructive guarantee on the existence of synthetic embeddings: a valid set of samples is always produced by design, unlike methods based on the adversarial generation of functional malware samples, which require a large corpus of real malware to be effective yet still cannot guarantee the creation of even a single successful evasion sample.

\subsection{Finetuning procedure}\label{sec:finetuning}

After generating the synthetic malware embeddings, we adapt the pretrained detector through a lightweight finetuning procedure. 
Let $\theta=\{\theta_f,\theta_h\}$ denote the parameters of the pretrained detection model, where $f$ denotes the feature extraction layers and $h$ the classifier head. During optimization, $\theta_f$ are frozen while only the parameters $\theta_h$ of the final classification layer are updated. This design preserves the latent representation learned from real binaries while refining the decision boundary only. At each training iteration, we sample a batch $B_r$ of real binaries $x$ and their corresponding labels $y$. Predictions are obtained from the pretrained detection model and optimized using the standard cross-entropy loss:

\[
\mathcal{L}_{\mathrm{real}}=
-\frac{1}{|B_r|}
\sum_{i=1}^{|B_r|}
\log p_{\theta}(y_i \mid x_i)
\]

We additionally sample a batch $B_s$ of synthetic embeddings $\tilde{\mathbf{m}}$ generated by the procedure described in Section~\ref{sec:embGen}
Since these embeddings are constructed to represent malware-like latent points, they are assigned the malware label $y=1$. The corresponding classification loss is

\[
\mathcal{L}_{\mathrm{syn}}=
-\frac{1}{|B_s|}
\sum_{j=1}^{|B_s|}
\log p_{\theta_h}(y_j=1 \mid \tilde{\mathbf{m}}_j)
\]

where $p_{\theta_h}(y_j=1 \mid \tilde{\mathbf{m}}_j)$ denotes the probability assigned by the classifier head $h$ to the malware class for the synthetic embedding $\tilde{\mathbf{m}}_j$. 
Note that, while the real binaries in $B_r$ are processed through the entire detector pipeline, including both the feature extractor $f$ and the classifier head $h$, each synthetic embedding $\tilde{\mathbf{m}}_j$ is processed only by $h$, as it already resides in the detector's embedding space.
The overall training objective is defined as

\[
\mathcal{L}=
\mathcal{L}_{\mathrm{real}}
+
\lambda\mathcal{L}_{\mathrm{syn}}
\]

where $\lambda>1$ is a confidence term that maximizes the malware probability assigned to generated representations. Since synthetic embeddings are designed to represent challenging malware-like points, increasing the confidence of their classification encourages the classifier to move the decision boundary away from benign regions of the latent space.

By jointly optimizing on real binaries and synthetic malware embeddings, the classifier learns to expand the malware region of the latent space around the benign manifold, thereby improving robustness against evading malware while preserving performance on known samples.

\section{Experiments}\label{sec:experiments}

In this section we present an extensive experimental evaluation of {\tt Phantomware} and compare it with state-of-the-art adversarial augmentation approaches. To this end, we first assess the effectiveness of a machine learning-based malware detector in the presence of emerging malware campaigns. Then, we apply the two augmentation strategies and compare the detection performance and computational efficiency of the resulting finetuned models. Finally, we evaluate the robustness of {\tt Phantomware} against adversarially crafted malware, and present an ablation study to quantify the contribution of its individual components to the overall detection performance.

We restrict our evaluation to Windows Portable Executable (PE) files, which represent the primary target of modern malware detection systems due to their widespread use in Windows environments. All experiments are conducted on real-world PE malware and goodware samples.

\subsection{Dataset}\label{sec:dataset}
For our evaluation, we use the BODMAS dataset \cite{yang2021bodmas}, which contains 57293 PE malware samples, including viruses, worms, and trojans, collected from multiple sources between August 2019 and September 2020. The dataset comprises 582 malware families, however many of these families contain only a small number of samples, making them unsuitable for our experimental setting. Thus, to ensure sufficient representation of each family, we retain only those with at least 100 samples. This filtering step results in a dataset containing 57 malware families and a total of 51974 malware samples. 

Goodware samples are collected from the PE Malware Machine Learning Dataset\footnote{https://practicalsecurityanalytics.com/pe-malware-machine-learning-dataset/}, composed of 86812 legitimate files extracted from instances of various versions of Windows 7 and above. To avoid an excessive imbalance between malware and benign samples, we randomly select 50\% of the available goodware instances for inclusion in our experiments.

\subsection{MalConv vs small malware families}

We perform a preliminary experiment to assess the capability of a machine learning-based detector to identify emerging malware campaigns. To this end, we employ the MalConv architecture proposed by Raff et al. \cite{raff2017malware}, specifically using the implementation provided in \cite{raff2021classifying}, which is memory use invariant to the length of the input, allowing us to train on files up to 16MB. 
We simulate an emerging malware campaign by selecting a target family and restricting its training data to the five samples immediately preceding the 10 most recent samples, which are reserved for testing. All earlier samples from the target family are excluded from training. MalConv is then trained for 30 epochs using these five target-family samples together with all samples from the remaining families in the dataset. The trained detector is evaluated on a test set comprising the 10 most recent samples from each family, including the target family, representing the next wave of variants that the detector would encounter after deployment.
This procedure is repeated, with each family in the dataset serving as the target family in a rotating fashion. For each of the 57 family-specific experiments, the resulting training set contained, on average, approximately 44.5K malware samples. This ensures that, while the target family is represented by only five samples to simulate an early-stage malware campaign, the detector is still trained on a sufficiently large and diverse malware background composed of the remaining families. Timestamps correspond to the first-seen date on VirusTotal.\footnote{https://www.virustotal.com/gui/home/upload}

\begin{figure*}[t]
    \centering
    \includegraphics[width=\textwidth]{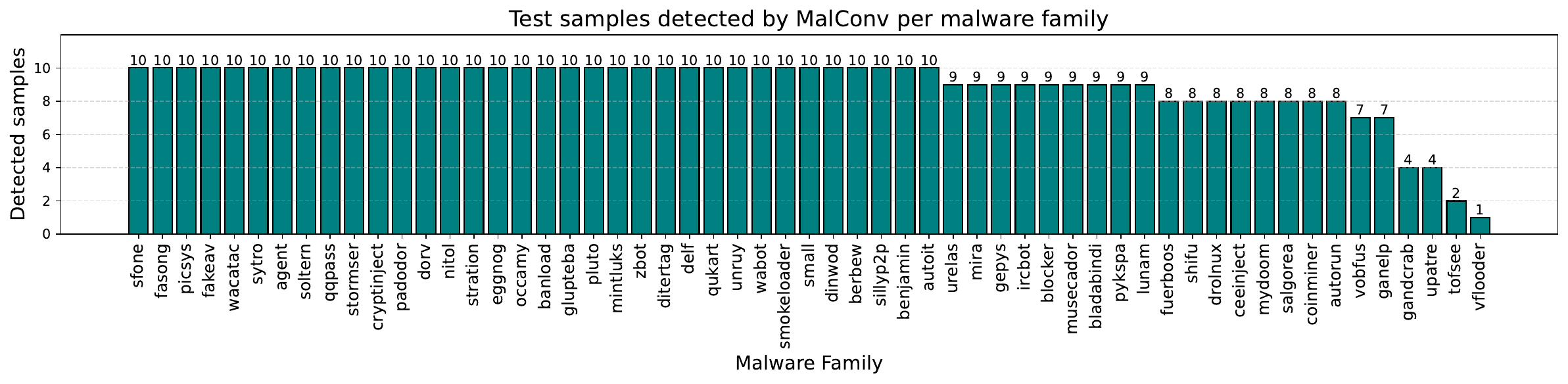}
    \caption{The detection rate achieved by MalConv on 10 test samples of 57 malware families.}
    \label{fig:malconvAccuracy}
\end{figure*}

The results, shown in Figure \ref{fig:malconvAccuracy}, show that cross-family generalization is highly inconsistent. Across 57 malware families, all variants are correctly detected for 34 target families (60\%), while MalConv fails to detect between 1 and 9 samples out of 10 of the remaining families.
Overall, MalConv misses 10.5\% of newly introduced variants. This variability highlights the limitations of relying solely on structural or behavioral similarities across families and underscores the need for data augmentation techniques that can support robust detection in scarce-data or early-stage malware campaign settings. 
The 23 families for which MalConv fails to correctly classify at least one of the ten test samples constitute the focus of our evaluation, as they correspond to scenarios in which the detector exhibits limited robustness against emerging threats.

\subsection{Evaluation of Phantomware}

To assess the effectiveness of {\tt Phantomware} in improving the detection of unseen variants from small malware families, we finetune MalConv using 50 synthetic malware representations for each target family. The goodware anchor set is selected as the union of 1000 boundary samples and 1000 frontier samples, while synthetic embeddings are generated by perturbing these anchors along the most malware-discriminative dimensions as detailed in Section \ref{sec:emb-generation}. All embeddings are represented as 128-dimensional vectors, corresponding to the latent representation produced by the adopted MalConv implementation \cite{raff2021classifying}.

\begin{figure}[t]
    \centering
\includegraphics[width=0.6\textwidth]{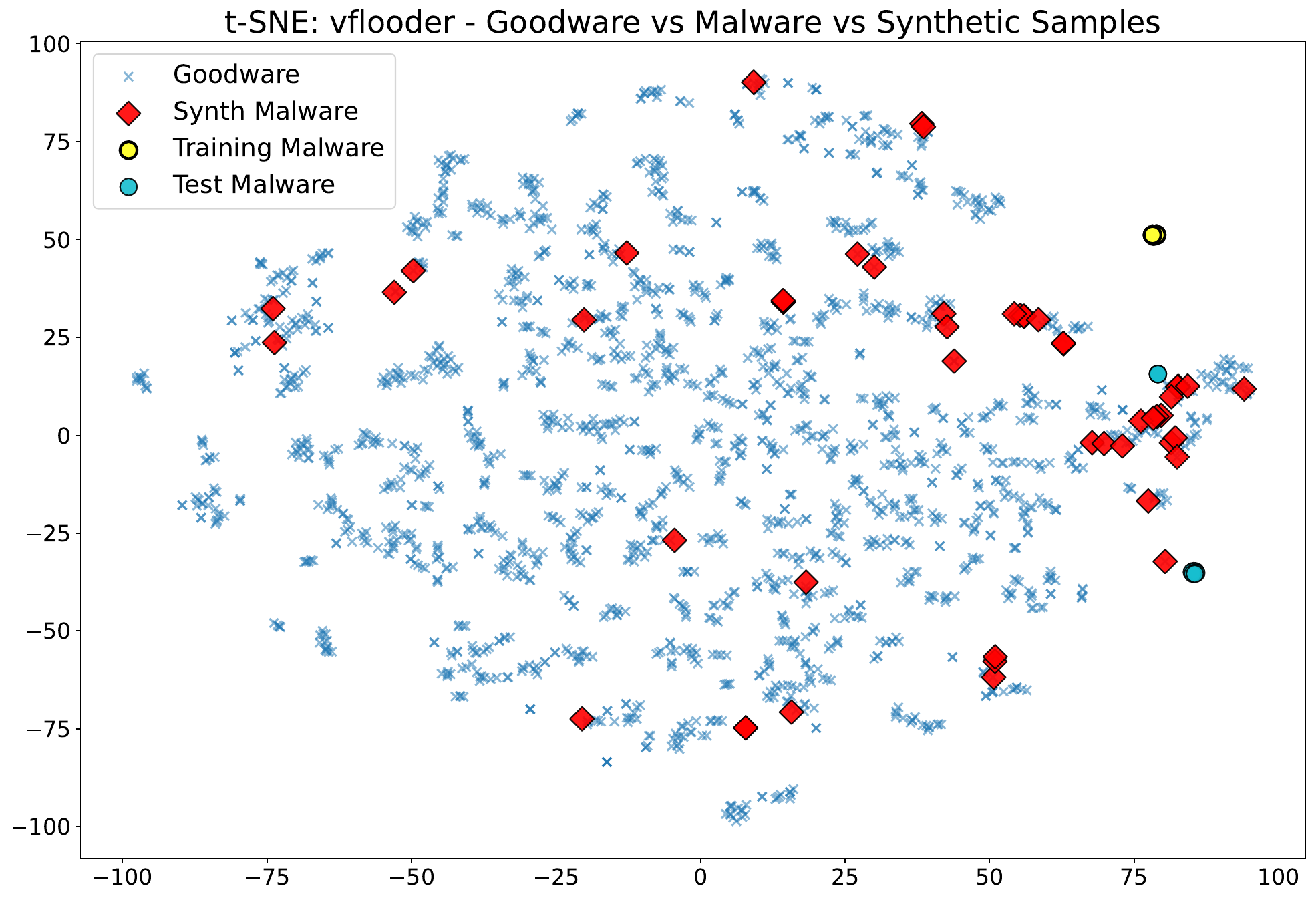}
    \caption{The 2D-tSNE projection of MalConv$_{Pha}$ embeddings for goodware samples, Vflooder malware samples, and Phantomware-generated synthetic samples.}
    \label{fig:tsneVflooderPhantomware}
\end{figure}

Figure \ref{fig:tsneVflooderPhantomware} shows the 2D t-SNE projection of the embeddings generated by {\tt Phantomware} in the MalConv latent space.
For visualization purposes, we report only the synthetic samples generated for the Vflooder malware family. We observe that the generated embeddings (red diamonds) do not only populate the vicinity of the original training samples (yellow circles), but span a broader region of the latent space, extending toward the areas occupied by the malware test samples (magenta circles), and frontier goodware samples as well (blue crosses).

The finetuning procedure is performed using the following hyperparameters: epochs=10, batch of real embedding samples $|B_r|=128$, batch of synthetic embedding samples $|B_s|=64$, $\lambda=2$.
Since finetuning the detector does not require the use of the entire original training set, we construct a representative subset of the available data that preserves the most informative characteristics of both classes. 500 goodware samples and 500 non-target family malware samples located close to the decision boundary are retained, as they correspond to regions where classification is most challenging.
To preserve the global structure of each class, we additionally perform clustering in the embedding space and select 500 representative samples associated with cluster centroids for both classes. The resulting finetuning set therefore combines 1000 hard samples, which provide information about the decision boundary, 1000 centroid representatives, which capture the overall distribution of each class, and 50 synthetic {\tt Phantomware}-generated samples.
We refer to the resulting finetuned model as MalConv$_{Pha}$. 

\begin{figure}[t]
    \centering
    \includegraphics[width=\textwidth]{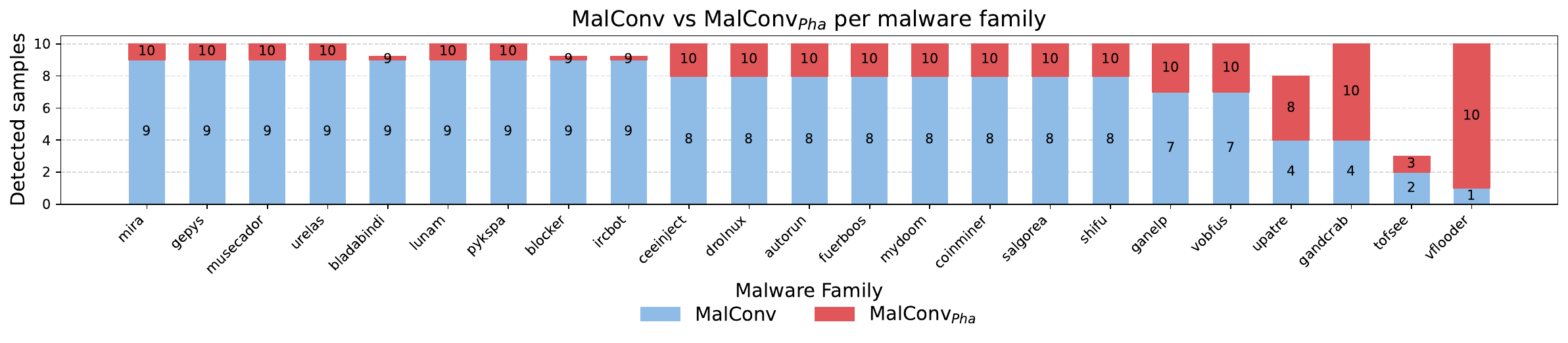}
    \caption{Impact of Phantomware augmentation on MalConv classification of target-family samples across malware families.}
    \label{fig:phantomAccuracy}
\end{figure}

Figure \ref{fig:phantomAccuracy} reports the detection rate achieved by MalConv$_{Pha}$ on the 10 test samples of each of the 23 target malware families. We observe that MalConv$_{Pha}$ improves the detection performance of the original MalConv model in 20 out of 23 cases, corresponding to 87\% of the considered families. Overall, MalConv correctly detects 510 out of 570 target-family test samples, achieving a detection rate of 0.89, whereas MalConv$_{Pha}$ detects 558 samples, increasing the detection rate to 0.98.

\begin{figure*}[t]
    \centering
    \includegraphics[width=\textwidth]{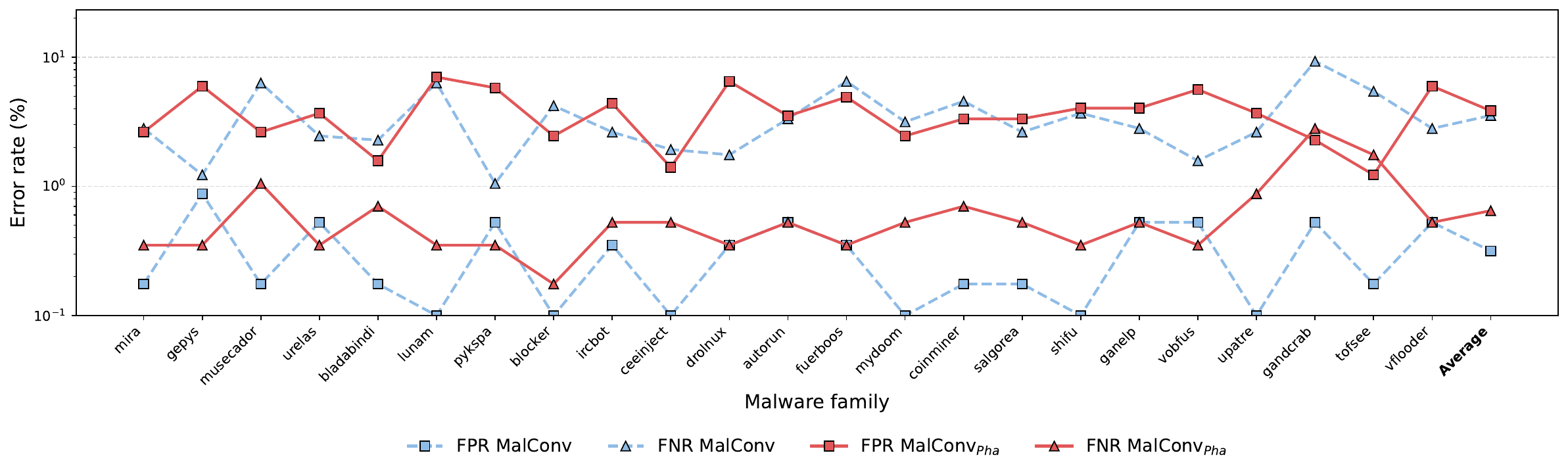}
    \caption{Misclassification rate of goodware (FPR) and malware (FNR) for MalConv (dashed-blue lines) and MalConv$_{Pha}$ (solid-red lines).}
    \label{fig:PhantomwareMisclassifications}
\end{figure*}

Figure \ref{fig:PhantomwareMisclassifications} illustrates the detection performance of MalConv$_{Pha}$ across the complete test set. Unlike the pretrained MalConv model, where the False Negative Rate (FNR) exceeds the False Positive Rate (FPR) by one order of magnitude, MalConv$_{Pha}$ exhibits an inverted trend: it detects a greater volume of malware, but this comes at the cost of missing more goodware. Specifically, while MalConv maintains an FPR consistently below 1\%, shifting the decision boundary closer to the benign regions of the latent space causes MalConv$_{Pha}$ to misclassify a larger fraction of benign samples, yielding an FPR between 1\% and 8\%. Notably, the FNR of MalConv$_{Pha}$ remains consistently lower than its FPR, indicating that the improvement extends beyond the target family to also include non-target-family malware samples that were previously misclassified by MalConv.
Notice that this trade-off between FNR and FPR is not unique to {\tt Phantomware}, but rather reflects a fundamental behavior of machine learning classifiers when the decision boundary is shifted toward the opposing class.

We repeat the same experiment by including in the finetuning set 500 synthetic malware embeddings instead of 50. The resulting performance is comparable to that reported in Figure~\ref{fig:phantomAccuracy}, suggesting that {\em diversity} is more important than {\em quantity}. Indeed, the initial set of 50 synthetic embeddings already provide sufficient diversity with respect to the original training samples, enabling MalConv$_{Pha}$ to adjust its decision boundary toward regions of the latent space occupied by unseen test samples. Increasing the number of synthetic embeddings by one order of magnitude does not provide additional benefits, indicating that they are redundant for augmentation purposes. This result suggests that the effectiveness of {\tt Phantomware} stems from its ability to generate a compact set of diverse and strategically positioned representations that expand the coverage of the latent malware space.

\begin{figure}[t]
    \centering
\includegraphics[width=0.5\textwidth]{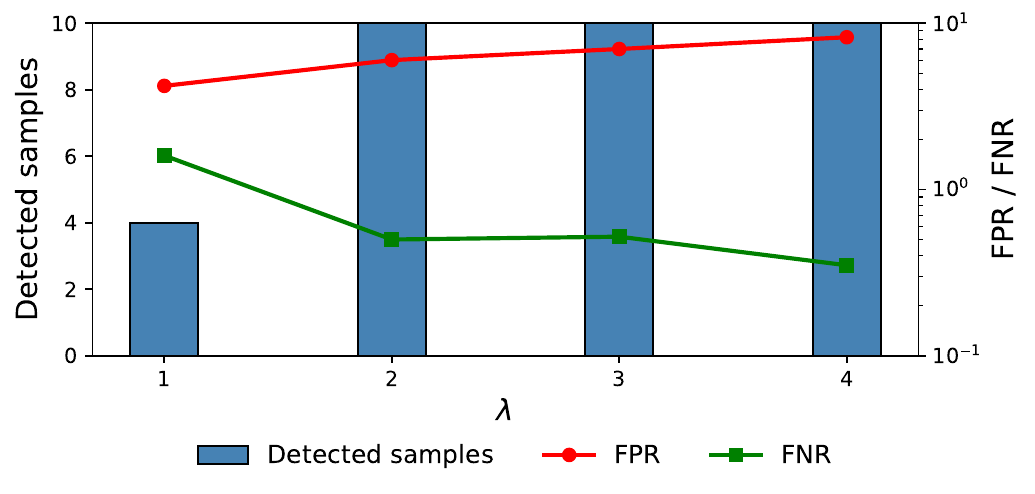}
    \caption{Impact of parameter $\lambda$ on the classification performance of MalConv$_{Pha}$ for the Vflooder family.}
    \label{fig:lambda-exp}
\end{figure}

Finally, we investigate the effect of the weighting parameter $\lambda$ on the classification performance. Figure~\ref{fig:lambda-exp} reports the results obtained on the Vflooder family by varying $\lambda$ in the range $[1,4]$. We focus on this family because it exhibits the largest performance gap between MalConv and MalConv$_{Pha}$, providing a wider margin for analyzing the impact of different $\lambda$ values. We first observe that values greater than 1 enable MalConv$_{Pha}$ to correctly detect all target-family test samples, validating our choice of assigning greater importance to the synthetic embedding objective $\mathcal{L}_{\mathrm{syn}}$. At the same time, increasing $\lambda$ also leads to a higher FPR and lower FNR on the entire test set, reflecting the expected trade-off between malware detection and benign-sample misclassification. This behavior is expected: as $\lambda$ increases, the optimizer places greater emphasis on correctly classifying synthetic embeddings as malware. Since these embeddings are intentionally generated close to the benign manifold, minimizing $\mathcal{L}_{\mathrm{syn}}$ encourages the classifier to expand the malware region toward nearby benign areas. Consequently, the decision boundary shifts toward the goodware distribution, improving malware recall at the cost of misclassifying a larger number of benign samples.

\subsection{Evaluation of adversarial approaches}\label{sec:adversarialExperiments}
In this section, we empirically validate the limitations of adversarial malware generation for data augmentation, showing that, consistently with the theoretical analysis presented in Section~\ref{sec:adversarialLimitations}, existing adversarial approaches fail to consistently produce sufficiently diverse and representative adversarial samples.

We generate adversarial malware samples using perturbation techniques that operate at the executable level. As discussed in Section~\ref{sec:RW}, adversarial malware generation consists of iteratively modifying a malicious executable while preserving its functionality, with the objective of reducing the detector's confidence until the sample is classified as benign, i.e., assigned a malicious probability below 0.5. Under the standard setting, this process yields a single adversarial example for each real malware.

To obtain a richer and more diverse approximation of the malware manifold, we extend this procedure beyond the first successful evasion. Rather than terminating the generation process as soon as one evading sample is found, we continue applying perturbations for up to $n$ iterations. This allows the attack algorithm to discover multiple distinct evading variants originating from the same real malware. As a result, we obtain a larger and more heterogeneous set of adversarial malware instances, providing a broader coverage of the regions of the feature space associated with successful evasion.

The adversarial generation techniques used in our experiments are implemented in SecML Malware \cite{demetrio2021secmlmalware}, an open-source Python framework for generating adversarial attacks against Windows PE malware detectors\footnote{https://github.com/pralab/secml\_malware}. The library provides both whitebox and blackbox attack strategies.
In the whitebox setting, the attacker has complete knowledge of the target detector, including its architecture, internal parameters, and confidence scores associated with each prediction, in addition to the final classification outcome. 
In the blackbox setting, by contrast, the attacker has no knowledge of the detector's internal structure and can only observe its classification output. 

In total, we use 11 adversarial generation techniques, divided into 6 whitebox and 5 blackbox.
The whitebox attacks are gradient-based adversarial evasion approaches, with an optimization process that proceeds iteratively: at each step, the attack computes the loss as the malware detection score; the gradients are projected onto the byte embedding space, and only the bytes with the largest gradient impact are updated by selecting the nearest valid byte embedding along the adversarial direction. This process is repeated until the maximum number of iterations is reached. The attacks are briefly described below:
\begin{itemize}
    \item {\bf Header\_perturbation}: randomly perturbs the bytes of the DOS header except for the MZ magic number and the pointer stored at offset 0x3c that are the only immutable fields required to preserve the validity of the executable \cite{demetrio2019explaining}.
    
    \item {\bf Header\_extension}: injects random adversarial perturbations into an additional space created within the executable by extending the DOS header, without altering the executable functionality \cite{demetrio2021adversarial}.

    \item {\bf HeaderFields\_perturbation}: randomly perturbs a subset of bytes in the DOS header and selected regions of the COFF/Optional PE headers, avoiding critical fields required for correct execution, while preserving malware functionality \cite{nisi2021lost}.

    \item {\bf Section\_shift}: injects adversarial random noise into a new space inside the executable created by shifting the content of the first section \cite{demetrio2021adversarial}.

    \item {\bf Padding\_injection}: appends random bytes at the end of the executable \cite{kolosnjaji2018adversarial}.

    \item  {\bf Slack\_perturbation}: randomly perturbes the slack bytes of an executable. Slack bytes are the unused or partially used bytes that exist due to alignment and file structure padding. They are typically not part of any meaningful PE structure and often contain leftover data \cite{suciu2019exploring}.

    \item {\bf Padding\_injection + Slack\_perturbation}: appends random bytes at the end of the executable, and randomly perturbs the slack bytes \cite{kreuk2018deceiving}.
\end{itemize}

The blackbox attacks are implementations of the GAMMA (Genetic Adversarial Machine learning Malware Attack) adversarial approach, proposed by Demetrio et al. \cite{demetrio2021functionality}. GAMMA is based on a genetic algorithm which tries to find the best modification by ranking and combining different sets of perturbations extracted from benign samples.
The attacks are briefly described below:
\begin{itemize}
    \item {\bf GAMMA\_API}: injects benign API into the PE binary import table.

    \item {\bf GAMMA\_header}: perturbs bytes not used by modern programs in the DOS Header.

    \item {\bf GAMMA\_headerFields}: perturbs selected bytes in the DOS Header, including altering section names, breaking the checksum, and altering debug information.

    \item {\bf GAMMA\_padding}: injects benign-like bytes at the end of the file.

    \item {\bf GAMMA\_sectionShift}: creates additional space before the beginning of a section, by shifting the content forward, and injects adversarial content in between.
\end{itemize}

To asses the applicability of the adversarial generation approach, we apply the above attack techniques to the 5 real malware samples from each target family, considering all the 57 families. Then we restrict the comparison study to the subset of 23 families for which MalConv fails to correctly classify at least one sample. We set the number of iterations to 15 and 80 for white-box and black-box attacks, respectively. These values were selected empirically as a practical trade-off between execution time and the number of successfully generated adversarial samples.

\begin{figure*}[t]
    \centering
    \includegraphics[width=\textwidth]{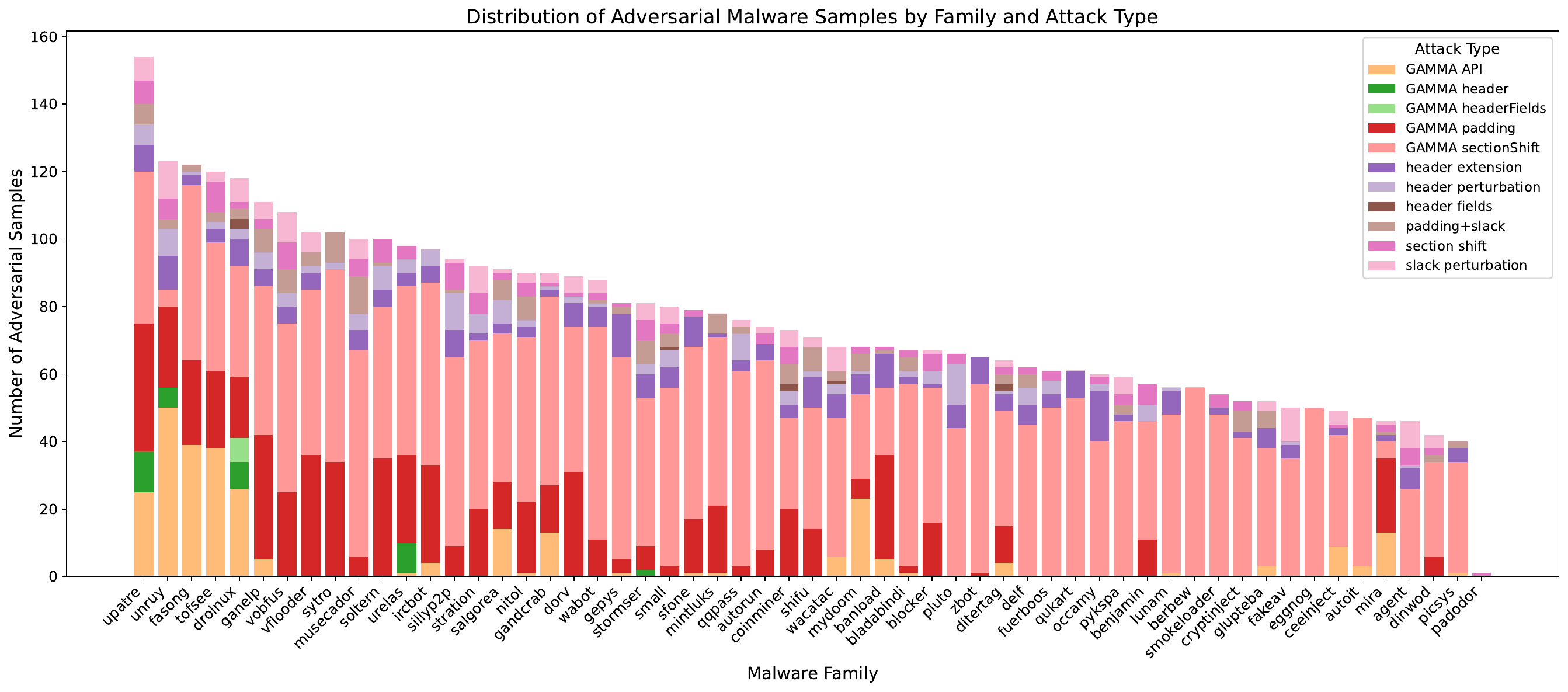}
    \caption{Distribution of malware adversarial samples grouped by family.}
    \label{fig:adversarialDistribution}
\end{figure*}

The statistics of generated adversarial examples are summarized in Figure \ref{fig:adversarialDistribution} and Table \ref{tab:adv_samples}.
As shown in Figure \ref{fig:adversarialDistribution}, the 11 attack techniques generate highly variable numbers of adversarial samples across malware families, collectively producing between 2 and 155 adversarial samples per family. 
This variability reflects the inherent limitation of adversarial generation algorithms we formally proved in Section \ref{sec:adversarialLimitations}, that is, they cannot guarantee the production of even a single successful evading sample within a finite number of iterations. 

The statistics reported in Table \ref{tab:adv_samples} further highlight this limitation. Only 3 out of the 11 attack techniques are able to generate at least one adversarial sample for {\em every} real malware sample (column "Adv/Real"), and the combination of all attacks produces an average of only 1.383 synthetic samples per real instance. Moreover, for most attacks, a large percentage of real malware samples do not lead to the generation of any adversarial variant (column "Real w/o Adv"), with GAMMA\_sectionShift being the only notable exception. This further confirms that the adversarial generation process does not consistently succeed across all malware samples. Finally, none of the considered attacks guarantees the generation of at least one adversarial sample for {\em every} malware family (column "Families w/o Adv"). Even GAMMA\_sectionShift, which achieves the broadest coverage among the evaluated attacks, fails to produce adversarial samples for one family (i.e., the Padodor family as shown in Figure \ref{fig:adversarialDistribution}). 

\begin{table*}[t]
\caption{Statistics of generated adversarial samples grouped by attack.}
\centering
\small
\setlength{\tabcolsep}{3pt}
\begin{tabular}{l|c|c|c|c}
\toprule
Attack & Adv Samples & Adv/Real & Real w/o Adv & Families w/o Adv \\
\midrule
Header extension & 272 & 0.954 & 44\% & 12\% \\
Header perturbation & 148 & 0.519 & 67\% & 33\% \\
HeaderFields perturbation & 9 & 0.032 & 98\% & 91\% \\
Padding+Slack & 144 & 0.527 & 71\% & 40\% \\
Slack perturbation & 147 & 0.516 & 80\% & 44\% \\
Section shift & 144 & 0.505 & 59\% & 30\% \\
GAMMA\_API & 288 & 1.010 & 79\% & 54\% \\
GAMMA\_header & 57 & 0.130 & 97\% & 91\% \\
GAMMA\_headerFields & 7 & 0.025 & 99\% & 98\% \\
GAMMA\_padding & 677 & 2.375 & 55\% & 33\% \\
GAMMA\_sectionShift & 2443 & 8.571 & 3.5\% & 1.8\% \\
\hline
All attacks & 4336 & 1.383 & 1.7\% & 0\% \\
\bottomrule
\end{tabular}
\label{tab:adv_samples}
\end{table*}

To evaluate the contribution of adversarial samples to the detection process, we finetune MalConv following the procedure described in Section \ref{sec:finetuning}: we use the feature extraction layers to compute the embeddings of the adversarial samples and subsequently use these embeddings to update the classifier head, while keeping the feature extractor frozen. We refer to the resulting model as MalConv$_{Adv}$. This finetuning procedure is performed independently for each of the 23 target malware families for which the original MalConv model misclassifies at least one sample.

To illustrate the unpredictable effect of adversarial samples on detection performance, we consider two representative extreme cases:

\begin{itemize}
    \item \textbf{Mira}, the malware family for which the 11 attack techniques collectively generate the fewest adversarial samples (46 in total), while the original MalConv model correctly detects almost all test samples (9 out of 10);

    \item \textbf{Vflooder}, one of the malware families for which the 11 attack techniques generate the largest number of adversarial samples (102 in total), while the original MalConv model correctly detects only a single test sample (1 out of 10).
\end{itemize}

\begin{table}[t]
\centering
\small
\caption{Analysis of MalConv$_{Adv}$ for Mira and Vflooder families.}
\label{tab:adversarial_analysis}
\begin{tabular}{l|cc|cc}
\hline
\multirow{2}{*}{\textbf{Attack}} & \multicolumn{2}{c|}{\textbf{Mira (9/10)}} & \multicolumn{2}{c}{\textbf{Vflooder (1/10)}} \\
\cline{2-3} \cline{4-5}
 & \textbf{\#Samples} & \textbf{Score} & \textbf{\#Samples} & \textbf{Score} \\
\hline
Header\_extension    & \cellcolor{gray!20}2  & \cellcolor{gray!20}9  & 5  & 10 \\
Padding+Slack       & \cellcolor{gray!20}1  & \cellcolor{gray!20}9  & \cellcolor{gray!20}4  & \cellcolor{gray!20}4  \\
Slack\_perturbation  & \cellcolor{gray!20}1  & \cellcolor{gray!20}9  & 6  & 10 \\
Header\_perturbation & 0 & -- & \cellcolor{gray!20}2  & \cellcolor{gray!20}4  \\
Section\_shift               & 2  & 10 & 0 & -- \\
GAMMA\_API           & \cellcolor{gray!20}13 & \cellcolor{gray!20}9  & 0 & -- \\
GAMMA\_padding       & \cellcolor{gray!20}22 & \cellcolor{gray!20}9  & 36 & 10 \\
GAMMA\_sectionShift      & 5  & 10 & \cellcolor{gray!20}49 & \cellcolor{gray!20}4  \\

\hline
combined attacks & \cellcolor{gray!20}39 & \cellcolor{gray!20}9 & \cellcolor{gray!20}55 & \cellcolor{gray!20}4 \\
\hline
\end{tabular}
\end{table}

Table \ref{tab:adversarial_analysis} reports the contribute of each attack technique $a$ as the number of test samples correctly detected by MalConv$_{Adv}^{a}$ (column "Score"), where MalConv$_{Adv}^{a}$ denotes the model finetuned exclusively on the adversarial samples produced by attack $a$. Column "\#Samples" indicates the number of adversarial samples generated by each attack.
The results confirm the behavior already observed with {\tt Phantomware}, that is, diversity appears to be more influential than quantity. Indeed, for both malware families, some attacks generate a very small number of adversarial samples (i.e., 2 samples with Section\_sift attack for Mira, and 5 samples with Header\_extension attack for Vflooder), yet these samples are sufficiently diverse from the original malware instances to enable MalConv$_{Adv}^a$ to correctly classify all test samples previously missed by MalConv. Conversely, other attacks produce substantially larger numbers of adversarial samples (i.e., 22 samples with GAMMA\_padding attack for Mira, and 49 samples with GAMMA\_sectionShift attack for Vflooder), but the generated samples lack sufficient diversity to achieve comparable improvements in detection performance. This finding further confirms the limitation discussed in Section~\ref{sec:adversarialLimitations}: the quality of the augmentation set generated by adversarial approaches is inherently unpredictable.

\begin{figure}[t]
    \centering
     \includegraphics[width=0.6\textwidth]{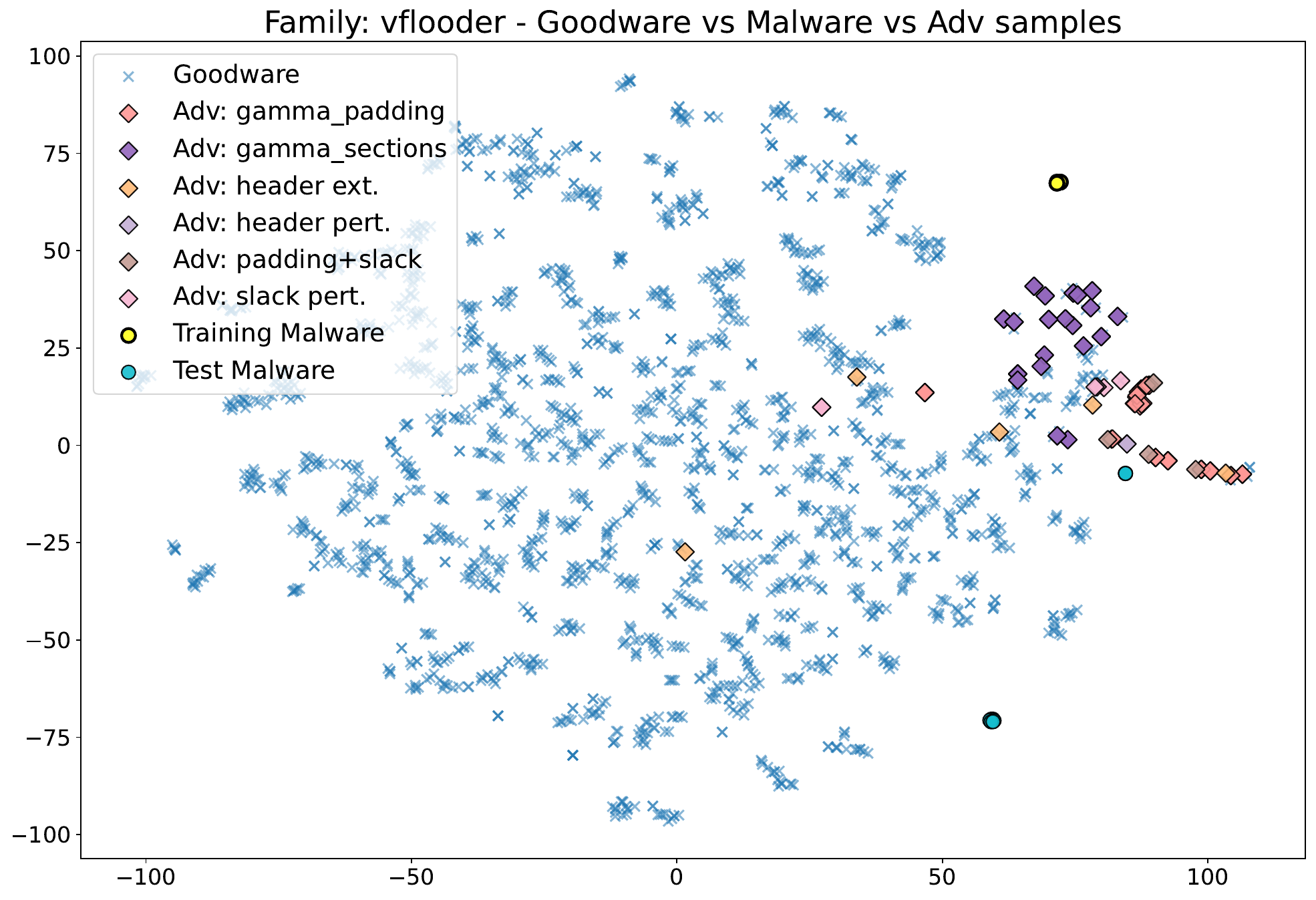}
    \caption{The 2D-tsne projection of Vflooder family's malware embeddings and of its adversarially generated samples.}
    \label{fig:tsneVflooderAdversarial}
\end{figure}

To better understand this phenomenon, Figure \ref{fig:tsneVflooderAdversarial} shows the 2D t-SNE projection of the embeddings produced by MalConv$_{Adv}$ for the Vflooder family. At first glance, most adversarial samples (colored diamonds) appear concentrated close to the original malware training samples (yellow circles). This behavior contrasts with {\tt Phantomware} (see Figure \ref{fig:tsneVflooderPhantomware}), which generates synthetic samples that span a broader region of the latent space, thereby providing a more diverse representation of potential malware variants.
A closer analysis reveals that adversarial samples generated with the Header\_Extension, Slack\_Perturbation, and GAMMA\_Padding attacks are located farther from the original malware training samples than those produced by the other attack strategies, introducing greater diversity into the augmented dataset. This explains why finetuning MalConv with these adversarial examples improves its ability to detect previously misclassified instances. Conversely, the GAMMA\_SectionShift attack produces a significantly larger number of adversarial samples (purple diamonds), however these samples occupy a restricted region of the latent space, introducing redundancy into the augmented dataset and providing limited additional information to the detector.

This dominance of diversity over quantity is further reinforced by the finding that finetuning MalConv with all adversarial samples from attacks that individually provided little or no  improvement does not yield any further gains. As shown in the "combined attacks" row of Table \ref{tab:adversarial_analysis}, aggregating samples from multiple attack techniques (grey cells) significantly increases the overall volume of training data, yet the resulting detection rate remains comparable to that achieved by the individual attacks alone. This suggests that simply increasing the number of adversarial samples does not compensate for their limited diversity.

A second important observation is that the same adversarial attack can exhibit markedly different effectiveness across malware families. For example, the GAMMA\_sectionShift attack enables MalConv$_{Adv}^{\gamma SecShf}$ to correctly detect all test samples of the Mira family using only five adversarial examples for finetuning. In contrast, for the Vflooder family, the same attack generates 49 adversarial examples but does not enable MalConv$_{Adv}^{\gamma SecShf}$ to detect all test samples. This result supports our initial hypothesis that the effectiveness of adversarial data augmentation is strongly dependent on the characteristics of the malware samples used to generate adversarial variants. 

\begin{figure*}[t]
    \centering
    \includegraphics[width=\textwidth]{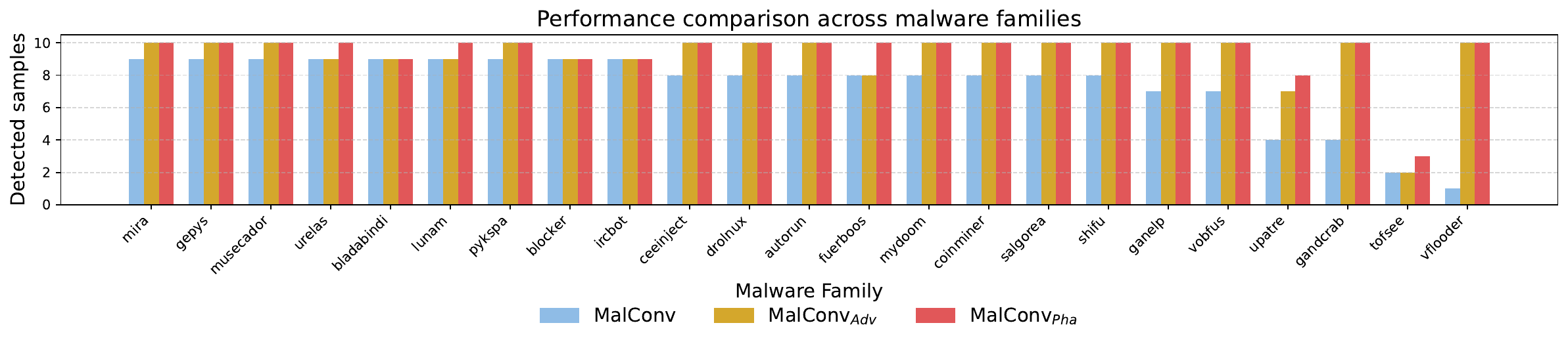}
    \caption{Detection rate of MalConv, MalConv$_{Adv}$, and MalConv$_{Pha}$ across malware families.}
    \label{fig:adversarialAccuracy}
\end{figure*}

Figure \ref{fig:adversarialAccuracy} reports the detection rate across the 10 test samples from each of the 23 target malware families achieved by MalConv$_{Adv}$ when finetuned with the adversarial samples generated by {\em all} attack techniques. Overall, MalConv$_{Adv}$ performs comparably to MalConv$_{Pha}$, except for five malware families (i.e., Urelas, Lunam, Fuerboos, Upatre, and Tofsee) for which MalConv$_{Pha}$ achieves superior results.

\begin{figure*}[t]
    \centering
    \includegraphics[width=\textwidth]{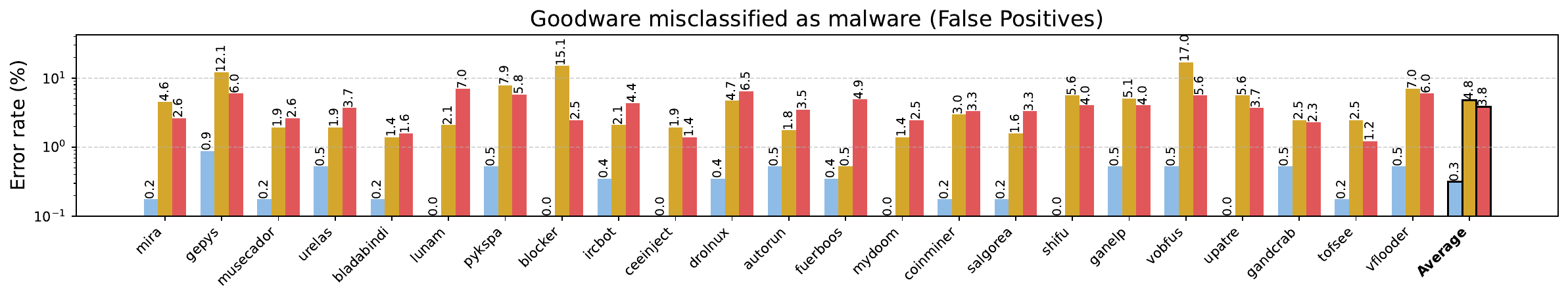}
    \includegraphics[width=\textwidth]{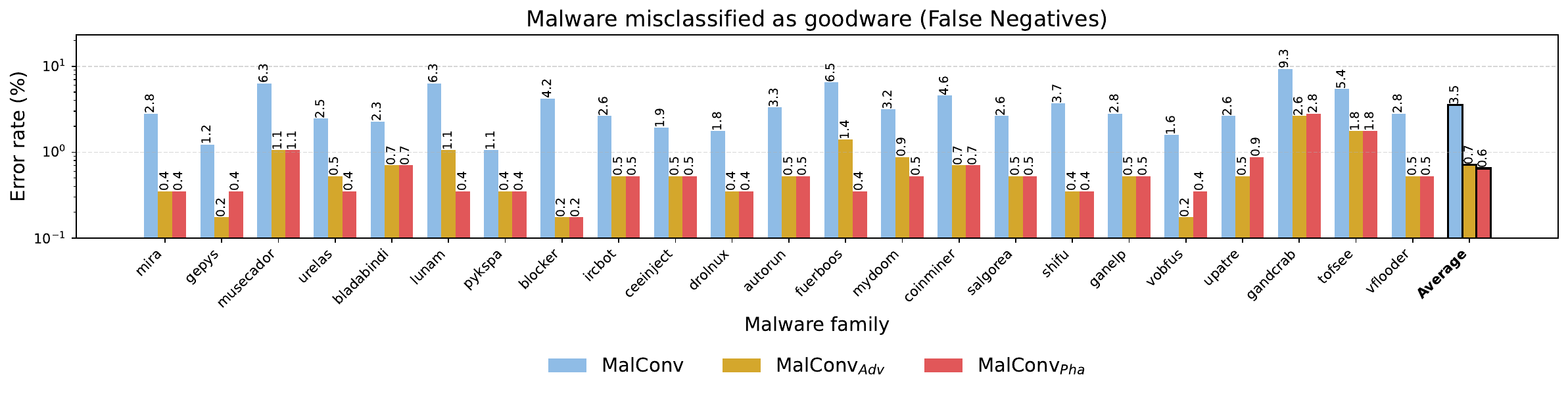}
    \caption{Misclassification rate of goodware (top) and malware (bottom).}
    \label{fig:misclassifications}
\end{figure*}

The detection performance of MalConv$_{Adv}$ on the complete test set is reported in Figure \ref{fig:misclassifications}. MalConv$_{Adv}$ exhibits behavior comparable to that of MalConv$_{Pha}$: the FNR is approximately one order of magnitude lower than that achieved by the pretrained MalConv model for most malware families, and this improvement comes at the cost of an increased FPR. This confirms that the trade-off between FPR and FNR is not specific to {\tt Phantomware} but is also observed with other data augmentation strategies.
On average, MalConv$_{Adv}$ performs slightly worse than MalConv$_{Pha}$ as shown by the rightmost series of Figure \ref{fig:misclassifications}. In particular, we note that its FPR exceeds 10\% for three malware families (i.e., Gepys, Blocker, and Vobfus).

\subsection{Execution times}

\begin{figure*}[t]
    \centering
\includegraphics[width=\textwidth]{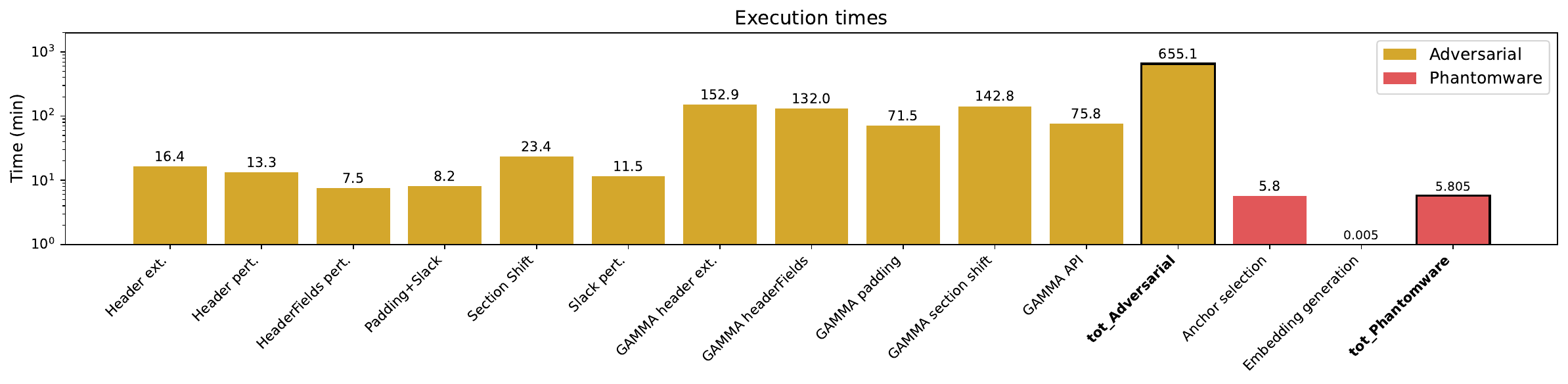}
    \caption{Execution time of adversarial approaches and Phantomware.}    
    \label{fig:executionTimes}
\end{figure*}

Figure \ref{fig:executionTimes} compares the average time required by the two approaches to generate augmentation data for a single malware family. 
For {\tt Phantomware}, it reports the time required by the two stages described in Section \ref{sec:embGen}, that are goodware anchor selection and synthetic embedding generation.
The results show that the time required by {\tt Phantomware} to generate synthetic embeddings is negligible compared with that needed for selecting the goodware anchors, which takes approximately six minutes in our experimental setting. This indicates that the computational cost of {\tt Phantomware} is dominated by the anchor selection phase and therefore scales primarily with the size of the goodware training set ($\sim$45K samples in our dataset).

For the adversarial approach, the figure reports the execution time of each of the 11 attack techniques used to generate adversarial malware samples, which ranges from 7.5 to 153 minutes, amounting to approximately 655 minutes ($\sim$7 hours) when all 11 attack techniques are executed. It is worth noting that these times account only for the generation of adversarial malware samples and do not include sample validation, which would require executing each generated variant in a controlled sandbox environment to verify that it preserves the malicious functionality of the original malware. Although this validation step is essential in offensive scenarios, it is unnecessary in our setting, where the generated samples are used solely for data augmentation in support of malware detection. 
Nevertheless, even without functional validation, adversarial malware generation remains substantially more time-consuming than {\tt Phantomware}. Including this additional step would further increase the overall execution time while potentially reducing both the number and the diversity of the resulting adversarial dataset, as variants failing the validation process would have to be discarded.

\subsection{Results discussion}
Our findings confirm that the effectiveness of data augmentation depends primarily on the diversity and representativeness of the generated samples rather than on their quantity. 
However, adversarial generation techniques provide no guarantees with respect to either of these properties, making the effectiveness of the resulting augmented dataset inherently unpredictable and at a considerably higher computational cost and execution time.
Successful adversarial augmentation requires that at least one of the adopted attack technique generates a sufficiently diverse and representative set of adversarial samples capable of guiding the detector toward a more appropriate decision boundary. While this condition is partially met in our experiments, achieving it required the use of 11 different attack techniques, with no guarantee that any of them would succeed. Thus, the observed improvements should be regarded as largely contingent on the specific attack–family combination rather than as a reliable property of adversarial augmentation. Indeed, our results show that no individual attack technique consistently generates sufficiently diverse and representative adversarial samples {\em across all target families}, highlighting the inherently unpredictable nature of adversarial generation as a data augmentation strategy.

In contrast, {\tt Phantomware} ensures the generation of a highly diverse and representative set of synthetic malware embeddings with minimal computational effort, enabling the finetuned detector to reshape its decision boundary toward previously unexplored regions of the latent space, and thereby improving its ability to generalize to unseen malware samples.

It is worth emphasizing that the adversarial augmentation results reported in this work are obtained using five real malware samples per target family. Since this amount of data already proves insufficient to consistently generate diverse adversarial samples with a single attack, further reducing the training set would likely exacerbate this limitation, making the resulting augmented dataset even less representative. In contrast, {\tt Phantomware} is specifically designed for extreme data-scarce scenarios, where the diversity of its synthetic embeddings is derived primarily from the goodware distribution rather than from the number of available malware samples, ensuring the generation of augmentation data even when only a single malware instance is available.

\subsection{Transfer attacks against Phantomware}

\begin{figure*}[t]
    \centering
    \includegraphics[width=\textwidth]{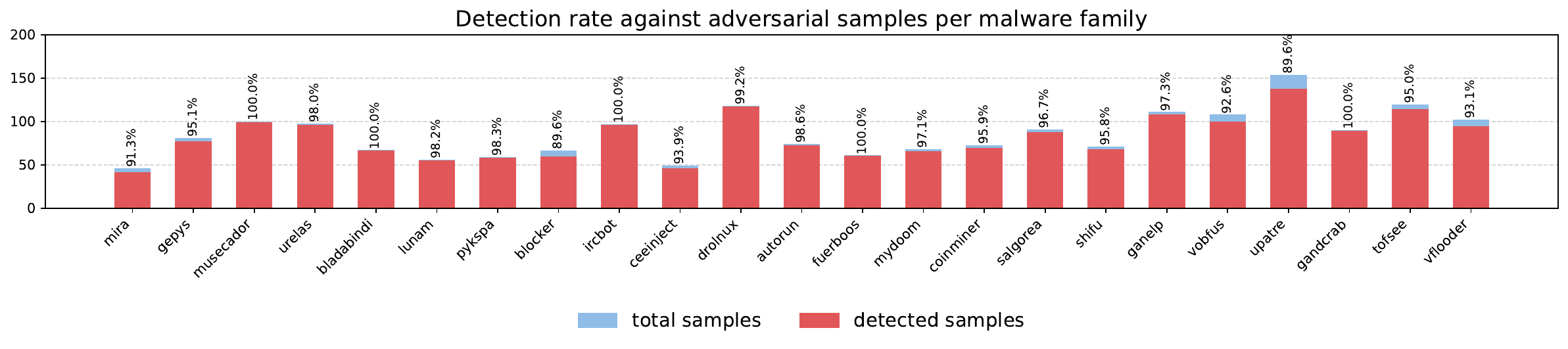}
    \caption{Detection rate of MalConv$_{Pha}$ against adversarial malware samples.}
    \label{fig:phantomwareVsAdv}
\end{figure*}

We next evaluate {\tt Phantomware} against {\em transfer attacks} \cite{demontis2019adversarial}. In a transfer attack, adversarial samples are optimized against a specific malware detector and then evaluated against a different detector. Such attacks are particularly practical in black-box settings, as attackers can craft adversarial samples using a classifier they control, without requiring access to or knowledge of the target model. Transfer attacks are more likely to succeed when the source and target detectors employ similar models and are trained on similar datasets. To emulate this setting, we evaluate the robustness of {\tt Phantomware} by testing MalConv$_{Pha}$ against adversarial samples crafted using MalConv as the source detector, following the procedure described in Section~\ref{sec:adversarialExperiments}.
Results reported in Figure~\ref{fig:phantomwareVsAdv} demonstrate that MalConv$_{Pha}$ maintains high detection performance against state-of-the-art adversarial techniques, successfully identifying 1886 out of 1961 samples for a 96\% detection rate. This result suggests that, by prioritizing structural diversity over learning specific evasion patterns, {\tt Phantomware} provides robust, generalized defense capabilities that effectively neutralize both emerging malware campaigns and adaptive adversarial threats.

In a recent study, Kozac et al.~\cite{kozak2025updating} proposed EXE-scanner, a malware detector trained exclusively on adversarial examples and designed to be chained with existing detectors. Its goal is to detect malware samples that evade the underlying detector, thereby mitigating the classification regressions that may arise when the existing model is fine-tuned or retrained.
The authors show that adversarial samples capable of bypassing the target detector are detected by EXE-scanner, increasing the detection rate from 91\% to over 97\%.
However, the adversarial examples used for evaluation are generated using the same attacks employed to construct the training set. Consequently, it remains unclear whether EXE-scanner would maintain its effectiveness against adversarial samples generated using attacks different from those adopted for the generation of the training set.
This raises concerns about the generalizability of the approach, as its effectiveness may heavily depend on the quality and diversity of the adversarial examples used for training. Since these samples are themselves generated using adversarial techniques, there is no guarantee that they adequately capture the diversity of {\em all} possible adversarial samples.

A related limitation is also evident in the adversarial-training strategy proposed by Li and Li \cite{li2020adversarial}. Their approach combines multiple attack algorithms and file manipulations to broaden the space of adversarial examples. The training procedure exploits them into a min-max optimization framework, in which the defender iteratively minimizes the model's loss on the strongest adversarial example of the current iteration, for a finite number of iterations, thereby approximating robustness against strongest attacks. While this strategy is effective for identifying and mitigating difficult attack instances, increasing the diversity of attack mechanisms and then optimizing for attack strength are not equivalent to explicitly maximizing the size, diversity, or coverage of a synthetic malware distribution. Indeed, the reported results indicate that greater attack diversity does not translate into robustness against broader and potentially unknown distributions of future malware variants.

In general, adversarial-training approaches treat adversarial examples primarily as a mechanism for exposing the detector to specific evasion behaviors. However, the number and diversity of adversarial samples are determined by the underlying malware corpus, the adopted attack mechanisms, and the ability of those attacks to successfully generate evasive variants. This limitation is particularly relevant in data-scarce settings, where a small malware corpus constrains the diversity of the resulting adversarial training set.
Also, it is impossible to predict whether future malware variants will arise from adversarial, syntax-level manipulations of existing samples or from non-adversarial, semantic evolutions of the target family. Consequently, an effective data augmentation strategy should generate synthetic samples that cover a broad spectrum of plausible malware variants rather than specializing in specific transformation mechanisms. Adversarial data augmentation inherently learns the perturbation patterns induced by the adopted attack techniques. As a result, when future variants are generated through different adversarial strategies, or evolve through entirely different semantic transformations, the learned robustness may not generalize effectively. 

{\tt Phantomware} addresses this limitation from a fundamentally different perspective. Rather than modeling specific evasion patterns, it enriches the latent malware distribution by generating diverse synthetic embeddings that span a broader region of the latent space. This enables the detector to learn more general malware characteristics, resulting in improved robustness against both adversarial and non-adversarial variants.

\subsection{Ablation study}
To better understand the contribution of the individual components of {\tt Phantomware}, we perform an ablation study focusing on the two main design choices of the proposed framework: the geometry-aware anchor selection, and the malware-discriminative perturbation direction. 

We evaluate the importance of the proposed goodware anchor selection strategy by replacing the boundary and frontier samples with a randomly selected subset of benign embeddings of equal size. This experiment assesses whether exploiting the geometry of the benign manifold is necessary for generating effective synthetic malware representations. Then, we separately analyze the contribution of the two anchor components by generating synthetic embeddings using only boundary samples or only frontier samples. This allows us to quantify the complementary role of near-boundary regions, which capture classification ambiguity, and peripheral benign regions, which provide coverage of unexplored areas of the latent space.

\begin{figure}[t]
    \centering
    \includegraphics[width=0.5\textwidth]{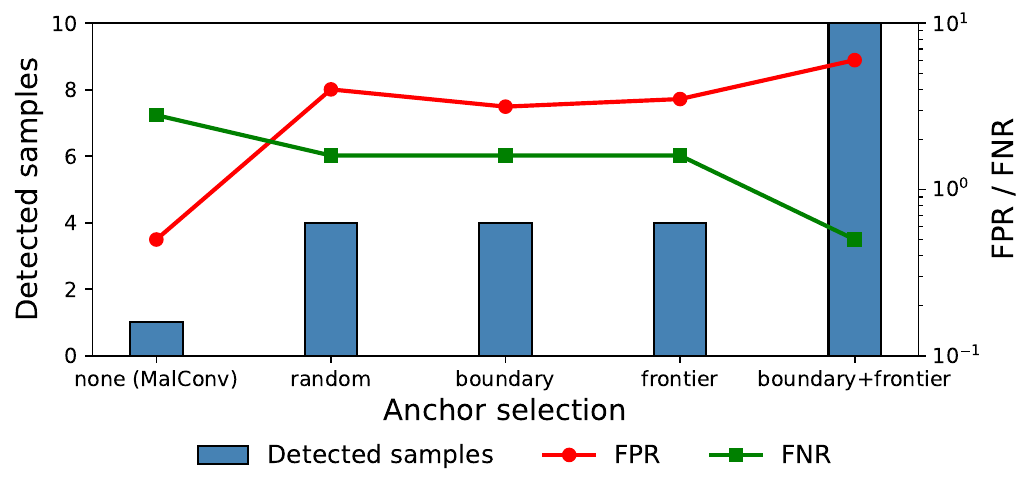}
    \caption{Ablation study: classification performance with four alternative anchor selection strategies.}
    \label{fig:ablation-anchor-selection}
\end{figure}

\begin{figure*}[t]
    \centering
    \includegraphics[width=\textwidth]{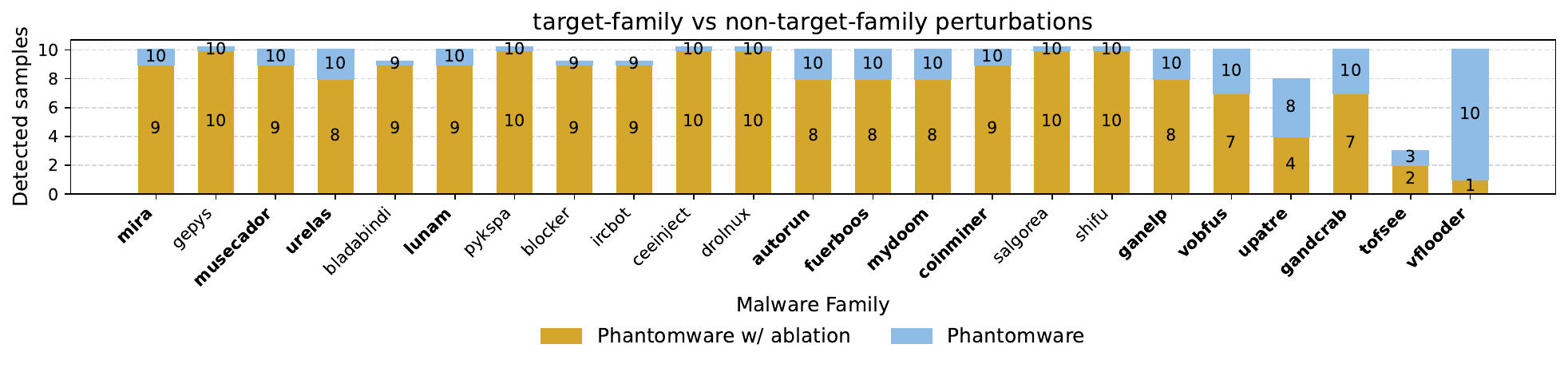}
    \caption{Ablation study: non-target-family vs target-family perturbations of goodware anchors. In bold are the families for which Phantomware performs better.}
    \label{fig:ablation-perturbation}
\end{figure*}

Figure~\ref{fig:ablation-anchor-selection} reports the classification performance obtained when selecting goodware anchors using four alternative strategies: random sampling, boundary samples only, frontier samples only, and the {\tt Phantomware} configuration, which combines both boundary and frontier samples. As in the previous experiments, we focus this ablation study on the Vflooder family, since it exhibits the largest performance gap between MalConv and MalConv$_{Pha}$, providing a wider margin for analyzing the contribution of the individual components of {\tt Phantomware}. We observe that each component provides complementary information: removing either the geometry-aware selection strategy or one of the two anchor subsets consistently reduces the detection rate of target-family samples (blue bars). Moreover, the original {\tt Phantomware} configuration achieves the lowest FNR (green line), as it also enables the detection of non-target-family samples that remain misclassified under the other configurations. This improvement, however, comes at the cost of a higher FPR (red line). Overall, these results confirm that explicitly exploiting the geometry of the benign manifold is crucial for generating effective and informative synthetic malware representations.

Finally, we investigate the importance of the malware-discriminative direction used during the generation process by replacing the target-family-guided perturbation with a direction extracted from a different malware family. This experiment assesses whether the effectiveness of {\tt Phantomware} depends on capturing latent characteristics specific to the target family, or whether generic malware-related directions are sufficient to generate effective synthetic representations.

Figure~\ref{fig:ablation-perturbation} reports the detection rates obtained when selected anchors are perturbed using either target-family-specific features (i.e., the {\tt Phantomware} configuration) or features extracted from non-target malware families. For the latter setting, a random family is selected for each target family. We observe that the use of target-family features provides a measurable advantage: in 14 out of 23 cases, perturbing anchors with target-family-specific directions achieves higher detection rates than using directions derived from unrelated malware families. This confirms that capturing the latent characteristics of the target family plays an important role in generating effective synthetic malware representations.

\section{Conclusions}\label{sec:conclusions}

In this paper we presented {\tt Phantomware}, a geometry-guided data augmentation pipeline designed to improve the robustness of pretrained malware detectors against unseen variants of small malware families. Through both formal analysis and empirical evaluation, we demonstrated the limitations of adversarial malware generation approaches for cybersecurity applications, particularly in data-scarce scenarios. These approaches typically require a substantial number of real malware samples, provide no guarantees on the diversity of the generated variants, and cannot even ensure the production of a single successful functional malware sample. In contrast, {\tt Phantomware} provides a practical alternative by operating directly in the latent space of a pretrained detector and guaranteeing the generation of synthetic augmentation data in a controlled manner, starting from only a handful of malware samples. Our experiments show that diversity, rather than quantity, is the key factor determining the effectiveness of augmentation data, and that latent-space generation enables the creation of challenging samples, more effective for downstream augmentation compared with problem-space approaches. These findings suggest the need for a shift in research direction, moving beyond purely offensive strategies focused on evasion capabilities toward data augmentation methodologies explicitly designed to improve the robustness and generalization of malware detectors.

As future work, we plan to extend {\tt Phantomware} to the Android malware domain, which presents similar challenges related to rapidly evolving threats and limited availability of samples from emerging malware families.

\bibliographystyle{plain}
\bibliography{bibliography}

\end{document}